# Cs adsorption on Bi$_2$Se$_3$

Haoshan Zhu, Weimin Zhou and Jory A. Yarmoff*

*Department of Physics and Astronomy, University of California, Riverside, Riverside CA 92521*

**Abstract**

Bi$_2$Se$_3$ is a topological insulator whose unique properties result from topological surface states (TSS) in the band gap. The adsorption of Cs onto a Bi$_2$Se$_3$ surface is investigated by low energy ion scattering and work function measurements. Much of the deposited Cs quickly diffuses to the step edges forming one-dimensional chains of positively charged adatoms, along with some deposition on the terraces. The work function decreases until a coverage of 0.1 ML is reached, beyond which it increases slightly. The minimum in the work function is due to depolarization of the dipoles induced when the concentration of adatoms in the chains reaches a critical value. A slow diffusion of adsorbed Cs from the terraces to the step edges is also marked by changes in the neutralization of scattered Na$^+$ and work function over time. The spatial distribution of the conductive charges in the TSS, which are primarily positioned between the first and second atomic layers, is confirmed by comparison of the neutralization of Na$^+$ scattered from Bi and Se.

*Corresponding author: Jory A. Yarmoff, e-mail: yarmoff@ucr.edu

**I. Introduction**

Topological insulator (TI) materials are characterized by topological surface states (TSS) that connect the conduction and valence bands [1,2]. The electrons in these TSS are responsible for the novel spin-dependent transport properties of TI materials and their potential utility in a variety of applications [2-4]. $Bi_2Se_3$, which is one of the most common TIs, is a two-dimensional material that consists of stacked quintuple layers (QL) ordered as Se-Bi-Se-Bi-Se. The QLs are bonded to each other through van der Waals (vdW) forces. The TSS of $Bi_2Se_3$ form a simple Dirac cone at the Γ point. First principles calculations have proposed that the TSS in $Bi_2Se_3$ are located almost completely within the outermost QL and that the spatial distribution of the TSS electrons is inhomogeneous [5,6].

Adsorption on TIs has been extensively studied due to the interest in these surfaces as electronic materials [7-9]. Although there is still a dispute about the effects of magnetic doping of TIs, nonmagnetic atoms or molecules adsorbed on TIs can alter the surface band structure but the TSS should be protected by time-reversal symmetry.

Alkali atom adsorption on TIs has also been studied extensively as a means to explore or modify the electronic structure [10-13]. Recent scanning tunneling microscopy (STM) and surface x-ray diffraction (SXRD) measurements, along with first principle calculations, show that most of the deposited Cs adatoms on $Bi_2Se_3$(0001) form one-dimensional chains along the upper portion of the step edges during the initial adsorption, while additional Cs adsorbs onto the terraces [14]. Ref. [14] also reports that Cs adsorption at Se vacancy sites is more energetically favorable than on intact areas of the terraces. Other work has indicated that there is a large energy penalty for large alkali atoms, such as Cs, to intercalate between the vdW gaps of TIs, although they are very mobile across the TI surfaces [12].

Low energy ion scattering (LEIS) is an experimental technique that has traditionally been used for surface elemental identification and atomic structural analysis of single crystal surfaces [15]. The penetration depth for low energy ions is less than a few atomic layers and, by choosing particular incident and emission angles, it is possible to probe only the outermost atomic layer [16,17]. In addition, the neutralization probability of scattered low energy alkali ions depends on the surface local electrostatic potential (LEP) above the scattering site [18-20]. This property of alkali LEIS has enabled investigations into the inhomogeneity of the LEP for single crystal surfaces that have spatial variations in the valence electron distribution [21] and for surfaces with submonolayer coverages of adsorbates [22-24]. Our previous work using neutralization in alkali LEIS successfully verified the inhomogeneous LEP and spatial distribution of the TSS states in $Bi_2Te_3$, which is a similar TI material to $Bi_2Se_3$ [25].

In the present work, LEIS and work function measurements are performed on $Bi_2Se_3$ surfaces exposed to Cs to investigate the adsorption sites and electronic properties. Analysis of the data confirms that Cs initially diffuses to the step edges, which leads to a concurrent reduction of the work function due to doping and the dipoles formed by the positively charged adatoms and their image charges. The work function has a minimum at a coverage of 0.1 monolayer (ML) when the Cs adatoms at the step edges are close enough to each other that their dipoles begin to depolarize. Samples monitored over long time periods indicate an additional slow, room temperature diffusion of Cs adatoms from the terraces to the step edges. Se surface vacancies are produced by $Ar^+$ ion bombardment, but there is no evidence of any preference for adsorption at these defect sites. In addition, the neutralization of $Na^+$ scattered from Cs-adsorbed surfaces confirms that the spatial distribution of the filled TSS in $Bi_2Se_3$ is consistent with calculations [5,6] and similar to that of $Bi_2Te_3$ [25,26].

**II. Experimental procedure**

Single crystals of $Bi_2Se_3$ [12] are grown using a slow-cooling method by melting high-purity Bi and Se shot following the recipe in Ref. [16]. An excess of Se is used to compensate for its high volatility that can lead to surface vacancies. The material cleaves easily along the (0001) plane producing samples around 10 mm in diameter. The samples are attached to a Ta sample holder by spot-welded Ta strips and cleaved in air several times to obtain a visually flat surface. They are then transferred into the main ultra-high vacuum (UHV) chamber through a load-lock.

Preparation and measurements of high-quality $Bi_2Se_3$ surfaces are performed inside the main UHV chamber, which has a base pressure of $2\times10^{-10}$ Torr. The surfaces are prepared by an $Ar^+$ ion bombardment and annealing (IBA) procedure described elsewhere [16,27]. It has been demonstrated that the samples prepared by IBA are QL-terminated with a surface quality equivalent to those cleaved in vacuum, but that they are also more resistant to surface contamination. The IBA procedure involves a preliminary degassing at 130°C for two hours, followed by a 2-hour bombardment by 0.5 keV $Ar^+$ with an average beam flux of about 200 nA $cm^{-2}$ and then a 30-min anneal at 130°C. After the preliminary process, the sample is ion bombarded for an additional 30 min and finally recrystallized by a 30 min anneal at 510°C. One cycle is generally sufficient to achieve a clean and ordered $Bi_2Se_3$ surface [27], but additional cycles are performed to ensure that the Ta sample holder is also clean, particularly if the work function of the holder is to be measured. Low energy electron diffraction (LEED) is employed to ensure the crystallinity of the $Bi_2Se_3$ surface.

Cs atoms are deposited onto the sample at room temperature by running 6 A through a well-outgassed SAES getter. The deposition rate is measured by LEIS to be around 0.02 ML $min^{-1}$ in the initial stages when sample surface is clean and the sticking coefficient should be near unity.

1 ML is defined as the number of first layer Se atoms on the $Bi_2Se_3$ crystal surface, which is 6.74x10$^{14}$ cm$^{-2}$. More than one hour of ion bombardment is needed to remove Cs from the surface after a large exposure, however, and Cs diffuses back to the surface, presumably from the sample holder, after 30 min of annealing at 510°C. Because of the difficulty in removing the Cs adatoms, the samples are not re-prepared by IBA before each Cs exposure, but instead a newly inserted sample is used only once with cumulative Cs depositions performed to collect a set of measurements.

Changes in the surface work function due to Cs adsorption are determined by bombarding the sample with a 200 eV electron beam and measuring the energy shift of the secondary electron cut-off via a modulation technique using the LEED optics [28]. The accuracy of the cutoff measurement is better than 0.1 eV, and the absolute value is calibrated by the known work function of the clean surface. The work functions are measured about 20 min after each Cs deposition, which is the fastest that this can be achieved due to positioning and setup time, but prior to any other measurements.

Time-of-flight (TOF) LEIS spectroscopy is performed using apparatus similar to that described in Ref. [22]. A pulsed beam of 3.0 keV Na$^+$ ions is produced from a thermionic emission gun. The scattered projectiles are collected by a triple microchannel plate (MCP) detector mounted at the end of a 0.57 m-long flight tube. The entrance to the MCP detector is held at ground potential to ensure equal sensitivity to charged and neutral species. There is a pair of parallel plates located in the flight tube that can deflect the scattered ions so that only the scattered neutral species are collected. For all of the TOF-LEIS data collected here, the ion beam is normally incident onto the sample and the emitted projectiles are collected at a scattering angle of 125°.

## III. Results and Discussion

### A. Adsorption of Cs

Figure 1 shows the surface work functions of a $Bi_2Se_3$ sample and the Ta sample holder as a function of Cs exposure. The absolute value of the work function is calibrated by setting the work function of IBA-prepared $Bi_2Se_3$ to the value of 5.4 eV that was measured by photoelectron spectroscopy for $Bi_2Se_3$ cleaved under UHV [29]. The work function of the $Bi_2Se_3$ sample drops from 5.4 eV to a minimum of ~3.8 eV after a 5 min Cs exposure, but it then increases to about 4.4 eV after an approximately 20 min deposition time and stays nearly constant with additional exposure. The work function is found to be uniform within an approximately 5 mm radius around the sample center.

The work function of the sample holder drops monotonically from 4.7 eV to 2.1 eV with Cs exposure. The work functions of pure Ta and Cs metal are 4.25 eV and 2.14 eV, respectively [30]. The initial work function of the sample holder is a little above the reference value because it is partially oxidized. The final value is close to the work function of Cs metal suggesting that the sample holder becomes completely covered by a Cs film after a sufficient exposure.

To better understand the behavior of Cs adsorption on $Bi_2Se_3$, the inset to Fig. 1 shows the work function as a function of the coverage of Cs on the surface. The coverage is determined from analysis of LEIS spectra, as described below. The amount of Cs on surface saturates at around 0.2 ML, indicating that the sticking coefficient drops to zero so that Cs does not form a complete film on $Bi_2Se_3$. Note that this saturation coverage is a factor of two smaller than is reported in Ref. [14], which is not too dissimilar a value, but the discrepancy could also be partially due to the different ways in which coverages are calculated. The work function is at its minimum at a Cs coverage around 0.1 ML and it increases with coverage above that until saturation. The saturation of the work function at such a low coverage suggests that additional Cs does not intercalate into the vdW gaps between QLs, consistent with reports in the literature [12], as otherwise the work function would be expected to continue to decrease with exposure even as the number of Cs adatoms on the surface saturates.

A minimum in the work function vs. coverage is commonly observed for alkali adsorption on many materials [31-34]. For Cs, a minimum in the work function typically occurs in the range of 2 to $4\times10^{14}$ atoms cm$^{-2}$ [34,35], which would correspond to a coverage range of 0.3 to 0.6 ML for Cs/Bi$_2$Se$_3$. Adsorbed alkali adatoms donate their valence electron to the substrate becoming positively charged and thereby forming upward pointing local surface dipoles in conjunction with their image charges [36-38]. At low alkali coverages, the charged adatoms on most surfaces repel each other so that they disperse across the sample as individual adatoms that act as isolated, non-interacting dipoles that reduce the work function monotonically with coverage. As the coverage increases, however, the distance between adatoms decreases to a certain point, which is typically around 1 nm [34], at which the alkali adatoms are close enough to each other for the dipoles to interact and depolarize so that the individual dipole strengths reduce and the overall work function increases slightly.

For Cs on Bi$_2$Se$_3$, Ref. [14] shows that initially Cs primarily adsorbs on the upper step edges, forming one-dimensional structures. At higher coverages, more of the Cs adsorbs on the terraces. A close-up analysis of the STM image shown in Ref. [14] indicates an average distance between Cs adatoms along a single step edge of around 3 nm, which is larger than the distance at which depolarization normally begins, although the coverage associated with the image is not reported. Note that the 0.1 ML coverage at the minimum in the work function, which is $6.7\times10^{13}$ cm$^{-2}$, is much smaller than the typical Cs coverage on metal surfaces at the minimum. The unusually small Cs coverage at the work function minimum supports the idea that the depolarization and concurrent increase in the work function occurs as the adatoms get close to each other along the one-dimensional steps, instead of on the terraces as occurs on most material surfaces. Thus, an overall coverage of 0.1 ML places the maximum amount of Cs on the steps that can be accommodated without depolarization.

Clean $Bi_2Se_3$ samples prepared by IBA show a sharp hexagonal 1x1 LEED pattern [27]. The LEED pattern, in conjunction with LEIS measurements, is also used to locate the azimuthal orientations for performing LEIS [16]. As more Cs is deposited, the adatoms do not develop higher order LEED patterns, unlike Cs deposited on other materials [22,35,39-41], but instead the 1x1 pattern gets blurrier. The degradation of the LEED pattern is likely due to the preference for adsorbed Cs to diffuse to and agglomerate at the step edges, as these chains are not well ordered [14]. Since Cs is a strong electron scatterer, a non-uniform distribution of Cs would lead to a blurring of the LEED spots.

TOF-LEIS is used to measure the surface elemental composition, termination and adsorbate coverage [16]. In LEIS, it is possible to orient the incoming ion beam and/or the position of the detector along low index crystalline directions so that single scattering is only possible from certain atoms in the crystal structure [16,17]. For all of the data presented here, the incident beam is along the surface normal so that only the three outermost atomic layers of $Bi_2Se_3$[0001] are directly visible, as the atoms in the 4$^{th}$ and deeper layers are completely shadowed by the atoms in the first three layers. Double alignment spectra are collected from the $Bi_2Se_3$ surface by positioning the detector along the $(1\bar{2}10)$ azimuthal plane at an angle of 55° from the surface normal so that the projectiles scattered from 2$^{nd}$ layer Bi and 3$^{rd}$ layer Se atoms are completely blocked from reaching the detector by the first layer Se atoms [16,42]. This azimuthal orientation is illustrated in the inset to the left panel of Fig. 2. A single alignment orientation, in which the incoming beam is along the surface normal while the detector is not positioned along a low index direction, allows all three of the outermost atomic layers to be probed. Single alignment is achieved here by rotating the sample azimuthally by 30° from the double alignment orientation, as illustrated in the inset to the right panel of Fig. 2. Spectra collected from IBA-prepared $Bi_2Se_3$ in these two orientations are provided in Ref. [16], and they confirm that the clean surface is terminated by the Se associated with an intact QL.

Se surface vacancies are reported to be energetically favorable sites for Cs adsorption [14]. To create Se surface vacancies here, a clean $Bi_2Se_3$ sample is sputtered using 500 eV $Ar^+$ ions with a fluence of $1.8 \times 10^{14}$ ions $cm^{-2}$ followed by a light anneal at 130°C for 90 min.

Figure 2 shows TOF spectra collected in both alignments before and after Cs exposure of the sputtered surfaces. The spectra are normalized by the incident beam fluence so that their absolute intensities can be directly compared to each other. The thick lines show the total yield, which includes both scattered ionic and neutral projectiles, while the shaded areas show the projectiles that were neutralized during scattering. The x-axis is reversed as shorter flight times correspond to higher scattered kinetic energies. TOF spectra show a distinct single scattering peak (SSP) for each element on the surface atop a background of multiply scattered projectiles. After Cs exposure, the total yield spectra contain three SSPs that correspond to $Na^+$ scattering from Se, Cs and Bi. The total yield spectra are discussed here, as they are most useful in measuring surface coverages since the neutralization process does not need to be considered. The neutral spectra are discussed below in terms of the electronic properties of the surface. Note that the differential scattering cross section increases with the mass of the target atom and MCP detector has higher sensitivity to particles with higher energy, which causes the Bi SSP to be considerably larger than that for Se even when the surface coverages are the same. In addition, the peaks broaden due to inelastic losses that the projectiles experience as they travel through the material, which are primarily due to electron excitation [15]. Note that the longer flight time peaks appear to be wider than those at shorter flight times, but this is an artifact of plotting the data vs. flight time. Each of the SSPs actually have approximately the same width in units of energy.

The coverages of each element, including that of Cs which was used for the x-axis of the inset to Fig. 1, are calculated by normalizing the area of the relevant total yield SSP to the area of the total yield Bi SSP after subtracting the multiple scattering background, as described previously [22]. To calibrate the coverage, it is assumed that the area of the Bi SSP measured from the clean, unsputtered sample in single alignment (data not shown) corresponds to the 1 ML of Bi in the second atomic layer. The Bi SSP is used for this calibration as it is the largest peak in the spectra, which leads to higher accuracy. To obtain accurate coverages, the SSP area ratios are corrected to account for both the differences in the differential cross-sections for $Na^+$ scattered from each element and the MCP detection efficiencies for projectiles with different scattered energies. The cross sections are calculated using the Thomas-Fermi-Moliere potential with the screening length reduced by an empirical formula [15]. The efficiency of the MCP detector as a function of scattered projectile energy is estimated from the data of Ref. [43]. Both of these effects act to increase the efficiency of projectile collection at higher scattered energies, which is why the Bi SSP is the largest one in single alignment. The error limits are determined by assuming that they are purely statistical in nature and thus equal to the square root of the total number of counts under each SSP. The calculated cross sections also have an inherent uncertainty due to a lack of accurate potentials [44,45], but this would only add a small systematic error to the reported coverages that would not alter any conclusions of the present paper.

In the double alignment spectra in the left panel of Fig. 2, a small Bi SSP is present after the clean surface has been sputtered. For clean defect-free surfaces, the Bi SSP is not observed in double alignment as is shown in Ref. [16]. The calibrated area of the Bi SSP represents 0.08 ML, which means that there are about 8% Se vacancies that leave the underlying Bi atoms exposed. It is assumed that most of these are isolated single atomic vacancies.

The LEIS results suggest that Cs adsorption at Se vacancies is less likely than adsorption on step edges or terrace sites. After exposure of the sputtered surface, a Cs SSP corresponding to 0.07±0.01 ML is present in both alignments and the number of scattered neutrals increases due to Cs doping. The Bi SSP in double alignment does not measurably attenuate while the Bi SSP in single alignment decreases, which suggests Cs does not preferentially adsorb onto Se vacancy sites as otherwise the Cs adatoms would shadow the second layer Bi atoms in double alignment. This result does not agree with the DFT calculations of Ref. [14], which conclude that Se vacancy sites are at least 115 meV more energetically favorable than other sites. A possible reason for this discrepancy is that the energy savings is too small to lead to a difference in the surface site occupancy at room temperature. Furthermore, the preference of Cs adsorption at step edge sites over Se vacancies shows that these two kinds of defects behave very differently and are not equivalent.

## B. Diffusion of Cs

The work function of Cs/$Bi_2Se_3$ not only changes with Cs coverage, but also changes over time as the sample sits in vacuum at room temperature. Figure 3 shows the change of work function as a function of time after various Cs exposures. When the Cs coverage is less than 0.1 ML, the work function is fairly stable, but when the coverage is larger than 0.1 ML, it increases by about 0.2 eV before stabilizing. The increase generally occurs more quickly for the samples that have a larger initial Cs coverage. The increase of work function over time suggests that adsorbed Cs is mobile and diffuses from the terraces to the step edges where the adatoms become closer together and thus more depolarized which causes the LEP above the Cs sites to increase. That the work function change occurs more readily at larger coverages is likely caused by the fact that more Cs adatoms are located near step edges, even though a limited amount of diffusion may also occur at lower Cs coverages.

The neutralization of scattered low energy alkali ions provides a unique method for probing the surface local electrostatic potential (LEP) [22-24,46]. When an alkali-metal atomic particle is in the vicinity of a surface, its ionization level shifts towards the Fermi level of the solid due to interaction with its image charge, and it also broadens due to overlap of the projectile atomic level and surface wave functions [47]. In the resonant charge transfer (RCT) model, which is typically used to describe alkali-surface interactions, electrons tunnel between the ionization level and the solid when the alkali is close to the surface. In low energy alkali scattering, the charge exchange is a non-adiabatic process in which the neutralization probability is frozen in along the outgoing trajectory while the projectile is still within a few Å's of the surface because the scattering times are short with respect to the electron tunneling rates. The neutralization probability, or neutral fraction (NF), of the particles scattered from a particular element is determined by dividing the area of the SSP in the neutrals spectrum by the area of the corresponding SSP in the total yield spectrum [22]. The measured NF depends on the value of the ionization potential, the degree to which the level shifts and broadens near the surface, and the LEP, which is sometimes referred to as the local work function, at the "freezing point" just above the scattering site. In general, the NF increases when the LEP decreases, and vice versa. This is the reason that the ratio of the neutral to the total yield SSPs increase in Fig. 2 after Cs deposition lowers the work function. If the surface potential were homogeneous, for example in a binary metal alloy with a free electron gas, the NF in scattering from each element would be the same.

Differences in the NF for scattering from different surface sites indicate, however, that the surface has an inhomogeneous potential. The NF in scattering from an isolated alkali adatom on a metal surface, for example, is generally larger than the NF in scattering from the substrate atoms due to the upwards dipole at the adatom site that reduces the LEP [18,48,49]. This effect is most pronounced at low coverages, where the strength of the individual dipoles is large. The prior work demonstrated that neutralization in alkali LEIS is sensitive to the LEP on a very local scale, and that it is a particularly useful tool for imaging local dipoles on a surface with an inhomogeneous potential.

The changes in NF with coverage and over time provide further insight into the surface diffusion of Cs. Figure 4 shows the coverages and measured NFs of the SSPs of all three elements as a function of the initial Cs coverage on a vacancy-free surface. The solid lines are the data collected within one hour of Cs exposure. The Cs coverage increases monotonically, as expected, and the data collected immediately after deposition is, by definition, a straight line with a slope of one. Both the Bi and Se coverages decrease with Cs coverage, however, due to shadowing by the Cs adatoms. The Cs SSP has the highest NF at all coverages due to the low LEP at the adatom sites, similar to alkali LEIS from alkali adatoms on metal and semiconductor surfaces [48]. The NF of the Cs SSP does not change significantly with coverage because it is determined by the combined effects of the decreasing global work function and the attenuating dipole strength at higher Cs coverages. The Cs NF has a large statistical error due to the small size of the SSP, however, particularly at the smallest coverages. The Bi and Se SSP NFs increase with Cs coverage due to the n-doping effect caused by Cs adsorption, as is expected.

The points connected by the dashed lines in Fig. 4 are obtained by collecting the same spectra after the samples have sat in the UHV chamber overnight for about 9 hrs. It is seen that the Cs surface coverage does not change over time. The NFs of all the SSPs decrease and the work function increases when the Cs coverage is close to or above 0.1 ML, however, which is the coverage at which the work function is at a minimum. These changes are consistent with the notion that Cs atoms continue to diffuse across the surface from the terraces to the step edges where their adsorption is more energetically favorable. When the Cs coverage is less than 0.1 ML, much of the Cs is concentrated at the steps but the distance between Cs adatoms is larger than the distance at which depolarization begins to occur, so that the work function and NFs do not change over time. When the Cs coverage is larger, however, Cs adatoms from the terraces diffuse to the step edges and the increased concentration at the steps places the adatoms close enough to each other so that the dipoles start to interact, which increases the LEP and decreases the NFs.

Although surface contamination by residual gas in the UHV chamber could also contribute to the changes of work function and NF over time, as Cs adatoms are sites at which residual gasses often adsorb, that does not appear to be occurring. If contamination played a major role, then the Cs SSP should also show some decrease over time as the adatoms become covered by oxygen or other species, which is not the case here. Also, the LEED pattern (not shown) does not degrade after the sample has sat in the chamber overnight, further suggesting that contamination is not playing a role.

## C. TSS charge distribution

An important observation is that the Se SSP measured from an unsputtered surface following Cs exposure always has a larger NF than the Bi SSP. The difference in the NFs is presumably due to the inhomogeneous TSS charge distribution within the first QL of $Bi_2Se_3$, as predicted by DFT calculations in the literature [5,6,26]. These DFT results indicate that the filled TSS are located beneath the $1^{st}$ and $3^{rd}$ layer Se atoms and above the $2^{nd}$ layer Bi atoms, thereby forming upward dipoles at the Se sites and downward dipoles at the Bi sites. The dipoles formed by the TSS cause an increase in the Se SSP NF and a decrease in that for Bi. This same effect was observed for $Na^+$ scattered from clean $Bi_2Te_3$ [25], but it is not apparent for clean $Bi_2Se_3$ because it has a higher work function that causes the NFs to both be near zero so that any differences are not large enough to be measured above the noise level. After adsorption of Cs, however, the overall work function of the sample is reduced so that the NFs all increase to values that enables the differences in the Bi and Se SSP NFs to be clearly distinguished.

The fact that the NF differences are maintained in the presence of Cs adsorbates shows that the TSS are not perturbed by the Cs adsorbates and have the same spatial distributions as on clean $Bi_2Se_3$ and $Bi_2Te_3$ surfaces. The stability of the TSS is likely due to their protection by time-reversal symmetry, but may also be partially because much of the Cs agglomerates at the step edges and thus has little effect on the electronic properties of the TIs that are determined by the band structure at the terraces.

In addition, the Bi SSP NF from a pre-sputtered surface with a 0.07 ML Cs coverage, as measured from the data in Fig. 2, is 0.29±0.05 for double alignment and 0.143±0.006 for single alignment. The NF in single alignment is similar to the value measured from the unsputtered surface in Fig. 4, which suggests that Cs adsorbs in the same manner on the pre-sputtered and unsputtered surfaces. The similarity in the single alignment NFs for as-prepared and pre-sputtered surfaces supports the conclusion reached above that there is no preference for adsorption at vacancy defect sites. The Bi SSP collected in double alignment from the pre-sputtered samples originates solely from $2^{nd}$ layer atoms above which there are no Se atoms, so that TSS electrons are presumably absent at these sites. In contrast, the Bi SSP in single alignment includes $2^{nd}$ layer Bi that is positioned below surface Se atoms so that the NF is expected to be affected by the presence of the TSS. The difference in which sites are probed would thus cause the NF of the Bi SSP in double alignment to be higher than in single alignment, which is consistent with observations. Therefore, the smaller Bi SSP NF for a pre-sputtered surface in single alignment as compared to double alignment further confirms that the NF of the Bi SSP measured from defect-free surfaces is reduced by the dipoles formed by the TSS electrons positioned above the $2^{nd}$ layer Bi atoms.

## IV. Conclusions

Both well-ordered $Bi_2Se_3$ surfaces and pre-sputtered surfaces that contain Se vacancies are exposed to Cs. The data indicate that Cs adsorbs as positively charged adatoms that initially agglomerate at the step edges. Above a 0.1 ML coverage, the Cs at the step edges become close enough to each other that the dipoles associated with the charged adatoms depolarize. With a sufficient Cs coverage, the work function and NF change over time suggesting that Cs adsorbed on the terraces continually diffuses to the step edges where their adsorption is more energetically favorable. There is no evidence in the present data that Cs prefers to adsorb at Se vacancy sites rather than on intact areas of the terraces. Thus, the role of any Se vacancies on the terraces is much different than that of atoms near the step edges, showing that these kinds of defects are not equivalent. In addition, Cs exposure reduces the overall work function so as to reveal that the spatial distribution of TSS charge is consistent with what is calculated by DFT and observed for clean $Bi_2Te_3$. This result demonstrates that alkali contamination does not perturb the TSS and thus should not interfere with devices fabricated from these materials.

**V. Acknowledgements**

This material is based on work supported by, or in part by, the U.S. Army Research Laboratory and the U.S. Army Research Office under Grant No. 63852-PH-H, and the National Science Foundation under Grant No. CHE - 1611563.

# References

1. M. Z. Hasan and C. L. Kane, Rev. Mod. Phys. **82**, 3045 (2010).

2. X.-L. Qi and S.-C. Zhang, Rev. Mod. Phys. **83**, 1057 (2011).

3. C. Nayak, S. H. Simon, A. Stern, M. Freedman, and S. Das Sarma, Rev. Mod. Phys. **80**, 1083 (2008).

4. Z. Jiang, C.-Z. Chang, M. R. Masir, C. Tang, Y. Xu, J. S. Moodera, A. H. MacDonald, and J. Shi, Nat. Commun. **7**, 11458 (2016).

5. W. Zhang, R. Yu, H.-J. Zhang, X. Dai, and Z. Fang, New J. Phys. **12**, 065013 (2010).

6. H. Lin, T. Das, Y. Okada, M. C. Boyer, W. D. Wise, M. Tomasik, B. Zhen, E. W. Hudson, W. Zhou, V. Madhavan, C.-Y. Ren, H. Ikuta, and A. Bansil, Nano Lett. **13**, 1915 (2013).

7. H. M. Benia, C. Lin, K. Kern, and C. R. Ast, Phys. Rev. Lett. **107**, 177602 (2011).

8. T. Valla, Z.-H. Pan, D. Gardner, Y. S. Lee, and S. Chu, Phys. Rev. Lett. **108**, 117601 (2012).

9. H. Zhu, W. Zhou, and J. A. Yarmoff, J. Phys. Chem. C **122**, 16122 (2018).

10. C. Seibel, H. Maaß, M. Ohtaka, S. Fiedler, C. Jünger, C.-H. Min, H. Bentmann, K. Sakamoto, and F. Reinert, Phys. Rev. B **86**, 161105(R) (2012).

11. M. Bianchi, R. C. Hatch, Z. Li, P. Hofmann, F. Song, J. Mi, B. B. Iversen, Z. M. Abd El-Fattah, P. Löptien, L. Zhou, A. A. Khajetoorians, J. Wiebe, R. Wiesendanger, and J. W. Wells, ACS Nano **6**, 7009 (2012).

12. A. G. Ryabishchenkova, M. M. Otrokov, V. M. Kuznetsov, and E. V. Chulkov, J. Exp. Theor. Phys+. **121**, 465 (2015).

13. Z.-H. Zhu, G. Levy, B. Ludbrook, C. N. Veenstra, J. A. Rosen, R. Comin, D. Wong, P. Dosanjh, A. Ubaldini, P. Syers, N. P. Butch, J. Paglione, I. S. Elfimov, and A. Damascelli, Phys. Rev. Lett. **107**, 186405 (2011).

14. M. M. Otrokov, A. Ernst, K. Mohseni, H. Fulara, S. Roy, G. R. Castro, J. Rubio-Zuazo, A. G. Ryabishchenkova, K. A. Kokh, O. E. Tereshchenko, Z. S. Aliev, M. B. Babanly, E. V. Chulkov, H. L. Meyerheim, and S. S. P. Parkin, Phys. Rev. B **95**, 205429 (2017).

15. H. Niehus, W. Heiland, and E. Taglauer, Surf. Sci. Rep. **17**, 213 (1993).

16. W. Zhou, H. Zhu, and J. A. Yarmoff, Phys. Rev. B **94**, 195408 (2016).

17. F. W. Saris, Nucl. Instrum. Meth. Phys. Res. **194**, 625 (1982).

18. C. B. Weare, K. A. H. German, and J. A. Yarmoff, Phys. Rev. B **52**, 2066 (1995).

19. K. A. H. German, C. B. Weare, P. R. Varekamp, J. N. Andersen, and J. A. Yarmoff, Phys. Rev. Lett. **70**, 3510 (1993).

20. T. Kravchuk, Y. Bandourine, A. Hoffman, and V. A. Esaulov, Surf. Sci. **600**, L265 (2006).

21. M. Aono, C. Oshima, S. Zaima, S. Otani, and Y. Ishizawa, Jpn. J. Appl. Phys. **20**, L829 (1981).

22. C. B. Weare and J. A. Yarmoff, Surf. Sci. **348**, 359 (1996).

23. J. A. Yarmoff, Y. Yang, and Z. Sroubek, Phys. Rev. Lett. **91**, 086104 (2003).

24. J. P. Gauyacq and A. G. Borisov, J. Phys.: Condens. Matter **10**, 6585 (1998).

25. W. Zhou, H. Zhu, and J. A. Yarmoff, Phys. Rev. B **97**, 035413 (2018).

26. T. Hirahara, G. Bihlmayer, Y. Sakamoto, M. Yamada, H. Miyazaki, S.-i. Kimura, S. Blügel, and S. Hasegawa, Phys. Rev. Lett. **107**, 166801 (2011).

27. W. Zhou, H. Zhu, C. M. Valles, and J. A. Yarmoff, Surf. Sci. **662**, 67 (2017).

28. J. H. Fritz and C. A. Haque, Rev. Sci. Instrum. **44**, 394 (1973).

29. M. T. Edmonds, J. T. Hellerstedt, A. Tadich, A. Schenk, K. M. O'Donnell, J. Tosado, N. P. Butch, P. Syers, J. Paglione, and M. S. Fuhrer, J. Phys. Chem. C **118**, 20413 (2014).

30. H. B. Michaelson, J. Appl. Phys. **48**, 4729 (1977).

31. B. N. J. Persson and H. Ishida, Phys. Rev. B **42**, 3171 (1990).

32. H. P. Bonzel, Surf. Sci. Rep. **8**, 43 (1988).

33. N. D. Lang, Phys. Rev. B **4**, 4234 (1971).

34. S. H. Chou, J. Voss, I. Bargatin, A. Vojvodic, R. T. Howe, and F. Abild-Pedersen, J. Phys.: Condens. Matter **24**, 445007 (2012).

35. J. Hrbek, Surf. Sci. **164**, 139 (1985).

36. R. W. Gurney, Phys. Rev. **47**, 479 (1935).

37. T. Aruga and Y. Murata, Prog. Surf. Sci. **31**, 61 (1989).

38. J. Topping, Proc. R. Soc. A **114**, 67 (1927).

39. C. A. Papageorgopoulos and J. M. Chen, Surf. Sci. **39**, 283 (1973).

40. A. Hohlfeld, M. Sunjic, and K. Horn, J. Vac. Sci. Technol. A **5**, 679 (1987).

41. R. D. Diehl and R. McGrath, J. Phys.: Condens. Matter **9**, 951 (1997).

42. X. He, W. Zhou, Z. Y. Wang, Y. N. Zhang, J. Shi, R. Q. Wu, and J. A. Yarmoff, Phys. Rev. Lett. **110**, 156101 (2013).

43. J. Oberheide, P. Wilhelms, and M. Zimmer, Meas. Sci. Technol. **8**, 351 (1997).

44. D. Primetzhofer, S. N. Markin, D. V. Efrosinin, E. Steinbauer, R. Andrzejewski, and P. Bauer, Nucl. Instr. Meth. Phys. Res. B **269**, 1292 (2011).

45. D. Primetzhofer, S. N. Markin, M. Draxler, R. Beikler, E. Taglauer, and P. Bauer, Surf. Sci. **602**, 2921 (2008).

46. J. J. C. Geerlings, L. F. Tz. Kwakman, and J. Los, Surf. Sci. **184**, 305 (1987).

47. J. Los and J. J. C. Geerlings, Phys. Rep. **190**, 133 (1990).

48. J. A. Yarmoff and C. B. Weare, Nucl. Instr. Meth. Phys. Res. B **125**, 262 (1997).

49. L. Q. Jiang, Y. D. Li, and B. E. Koel, Phys. Rev. Lett. **70**, 2649 (1993).

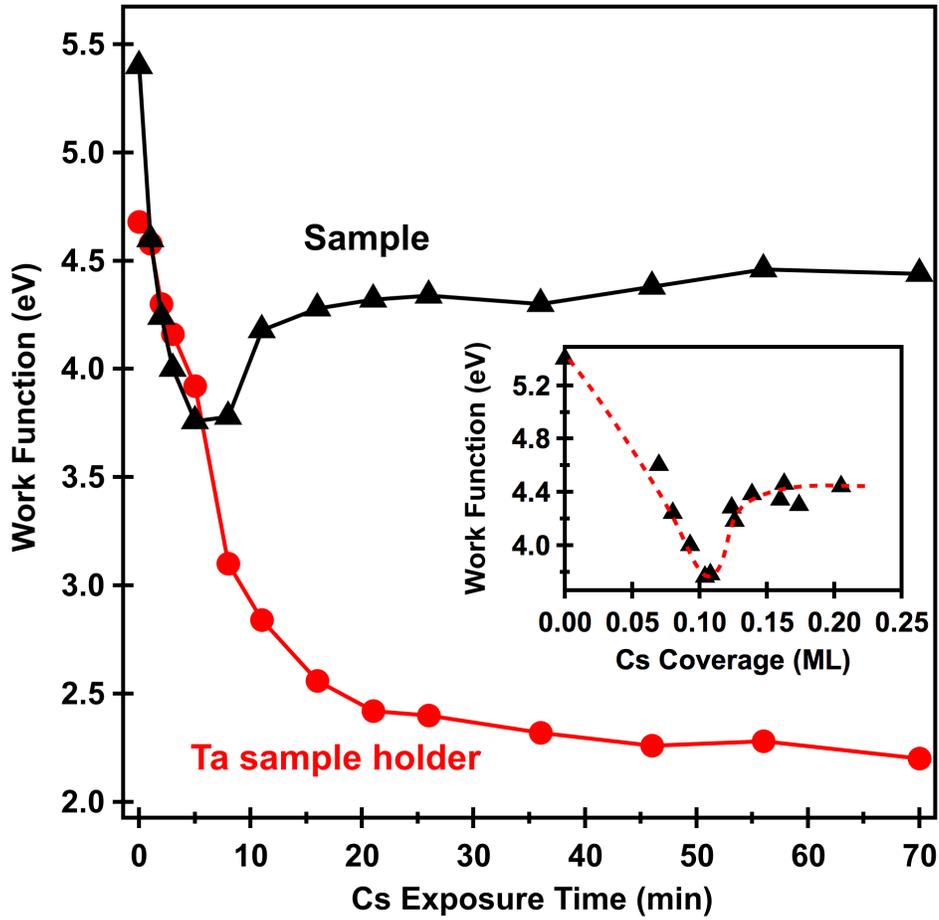

**Figure 1.** Work function of an IBA-prepared Bi$_2$Se$_3$ surface (solid triangles) and the Ta sample holder (solid circles) shown as a function of Cs exposure time. The inset is the same work function data, but instead given with respect to Cs coverage (see text). The dashed line illustrates the trend of the work function change as an aid to the eye and is not a fit to the data.

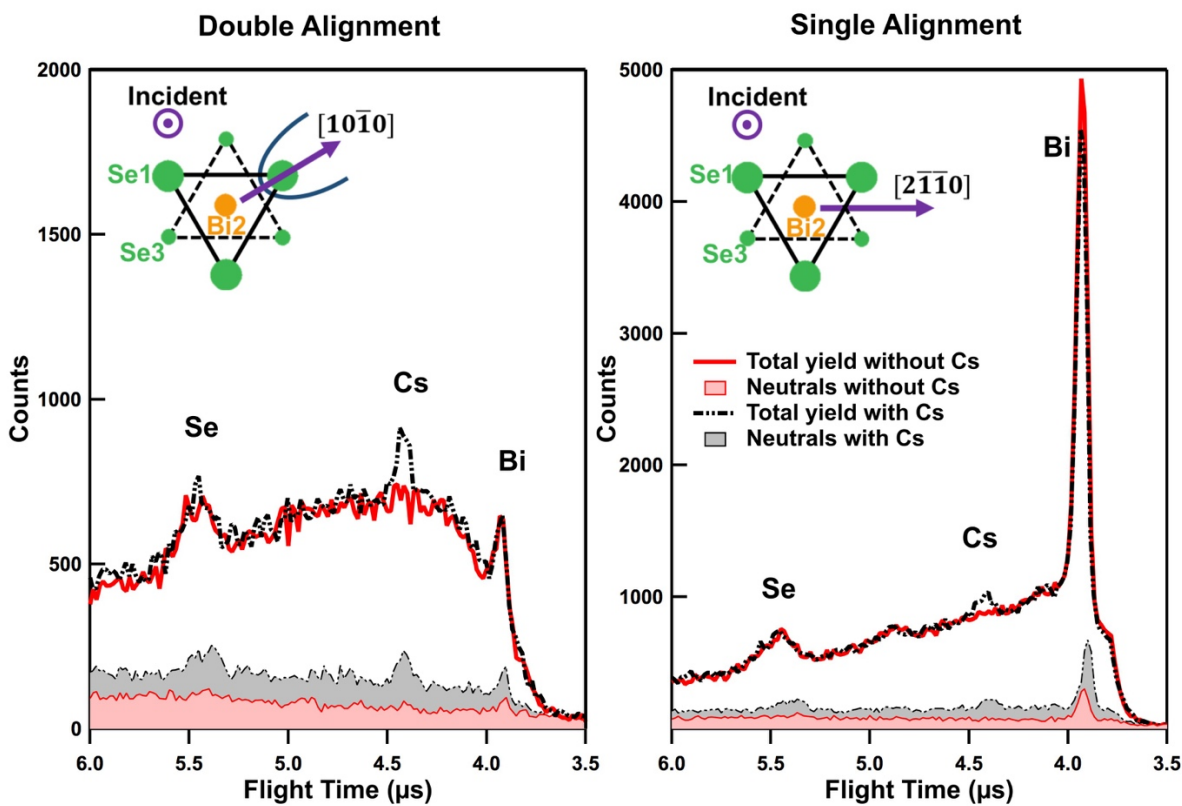

**Figure 2**. TOF spectra of the scattered total yield and neutrals for 3.0 keV Na$^+$ in both double (left panel) and single (right panel) alignments before and after 0.07 ML of Cs is deposited onto Bi$_2$Se$_3$ surfaces that contain about 8% Se vacancies created by Ar$^+$ sputtering. The Na$^+$ ion beam is incident normal to sample surface and the detector is positioned at a scattering angle of 125° along the indicated azimuths. The spectra are normalized to the incident beam fluence. The insets to each panel show a schematic top view diagram of the crystal structure in which the atoms and their layer numbers are indicated along with projections of the azimuthal directions of the scattered beams.

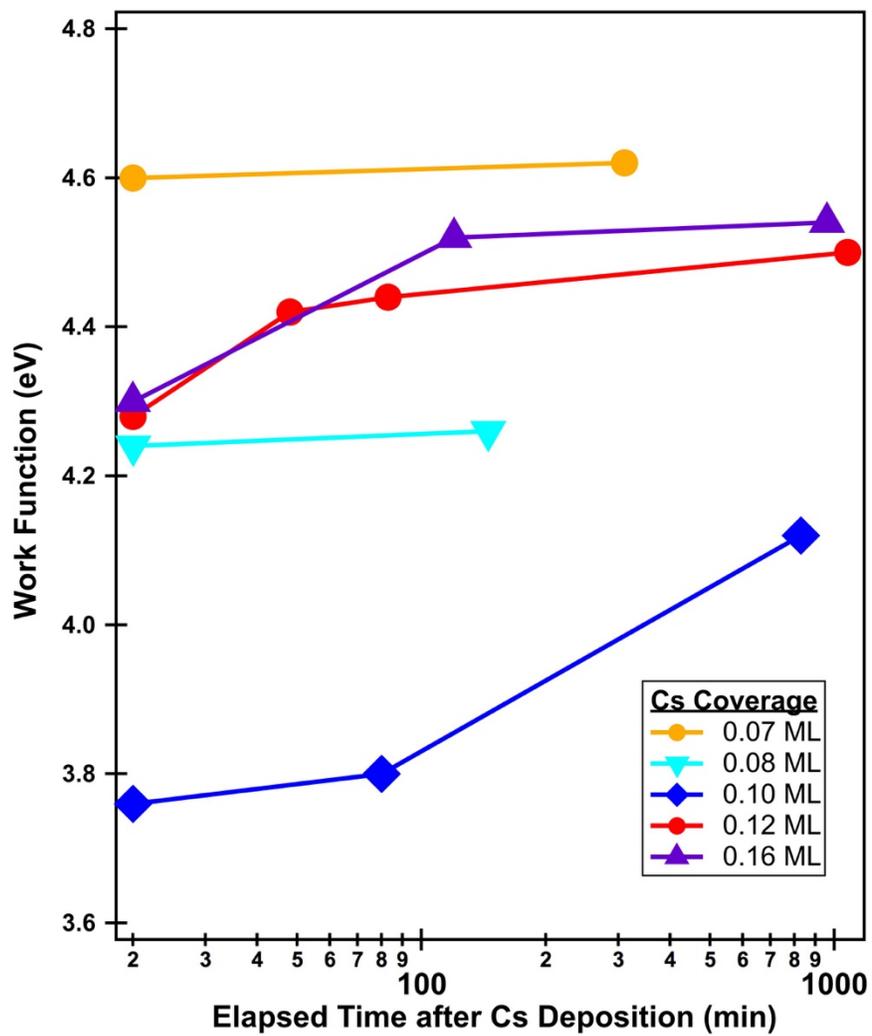

**Figure 3**. The measured work function with the samples held at room temperature shown as a function of time after various Cs depositions on IBA-prepared $Bi_2Se_3$.

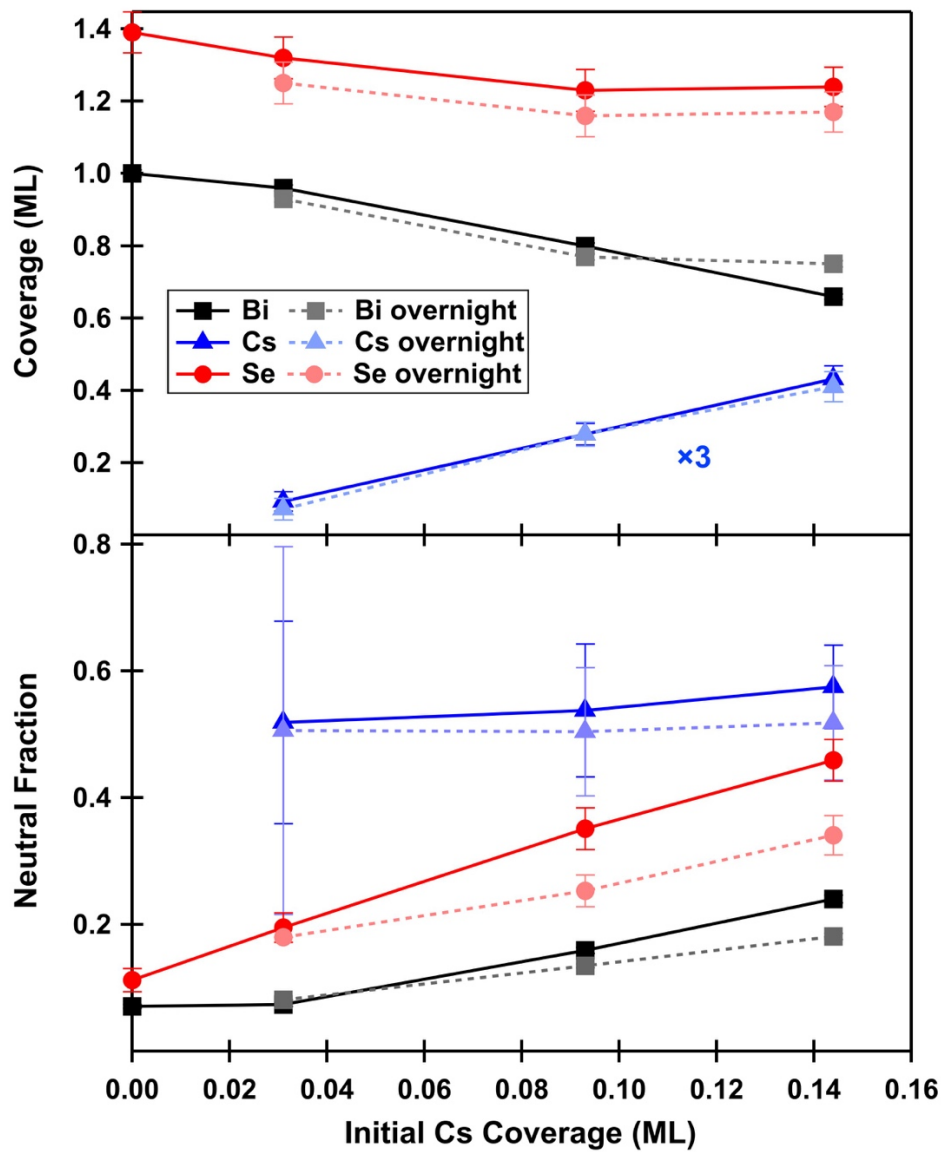

**Figure 4**. Se, Cs and Bi coverages and neutral fractions for 3.0 keV Na$^+$ singly scattered in single alignment from an IBA-prepared Bi$_2$Se$_3$ sample shown as a function of the initial Cs coverage. The points connected by solid lines were derived from LEIS spectra collected immediately after deposition, while the points connected by dashed lines are from data collected after the samples sat in the UHV chamber overnight (about 9 hrs) at room temperature.